\documentclass[12pt,a4paper]{iopart}

\usepackage[utf8]{inputenc}
\usepackage[british]{babel}
\usepackage{textcomp,graphicx,cite}

\begin{document}

\title[Fluctuations and differential contraction in Hydra regeneration]{Fluctuations
and differential contraction during regeneration of Hydra vulgaris tissue
toroids}
\author{Michael Krahe\textsuperscript{1}, Iris Wenzel\textsuperscript{1},
Kao-Nung Lin\textsuperscript{1}, Julia Fischer\textsuperscript{1}, Joseph
Goldmann\textsuperscript{2}, Markus Kästner\textsuperscript{2}, and Claus
Fütterer\textsuperscript{1,3
}}
\address{$^1$Fakultät für Physik und Geowissenschaften, Institut für
Experimentelle Physik I, Universität Leipzig, 04103 Leipzig, Germany}
\address{$^2$Fakultät Maschinenwesen, Institut für Festkörpermechanik,
Technische Universität Dresden, 01062 Dresden, Germany}
\address{$^3$Translationszentrum für Regenerative Medizin (TRM), Universität
Leipzig, 04103 Leipzig, Germany}
\ead{c.fuetterer@biophysik.net}

\vspace{16pt}
\begin{indented}
\item[]\rm\emph{We wish to dedicate the present publication to Malcolm Steinberg
($\dagger$ February 7, 2012).}
\end{indented}

\begin{abstract}
We studied regenerating bilayered tissue toroids
dissected from Hydra vulgaris polyps and relate our macroscopic observations
to the dynamics of force-generating mesoscopic cytoskeletal structures.
Tissue fragments undergo a specific toroid-spheroid folding process leading to complete
regeneration towards a new organism.
The time scale of folding is too fast for biochemical
signalling or morphogenetic gradients which forced us to assume purely 
mechanical self-organization.
The initial pattern selection dynamics was studied by embedding toroids into hydro-gels
allowing us to observe the deformation modes over longer periods of time.
We found increasing mechanical fluctuations which break the toroidal symmetry and
discuss the evolution of their power spectra for various gel stiffnesses.
Our observations are related to single cell studies which explain
the mechanical feasibility of the folding process.
In addition, we observed switching of cells from a tissue bound
to a migrating state after folding failure as well as in tissue injury.

We found a supra-cellular actin ring assembled along the toroid's inner edge.
Its contraction can lead to the observed folding dynamics as we could confirm by finite element 
simulations.
This actin ring in the inner cell layer is assembled by myosin-driven length
fluctuations of supra-cellular $\alpha$-actin structures (myonemes) in the outer
cell-layer. 
\end{abstract}


\section{Introduction}
Regeneration and growth of tissues have mainly been investigated on two scales,
the macroscopic one, where the tissue is considered
as a piece of continuous material, and the molecular one, where the tissue
dynamics
is reduced to biochemical signalling. The impressive recent results of cellular
and molecular biophysics, however, have revealed a surprizing complexity
of the cytoskeletal dynamics. The question, what this complexity is required for
may partially be answered by the living conditions in a
collective environment.
However, the findings about single cells have been integrated into the picture
only rudimentarily, so far. We try to close the gap and investigate physical
phenomena at a mesoscopic level by combining a minimum of sub-cellular and
molecular structures with a coarse-grained description, e. g. as a solid or
fluid, in order to explain our experimental findings. However, this field is
still in its very infancy and many questions remain to be investigated.

\begin{figure}[htp]
\centering
\includegraphics[height=4.5cm]{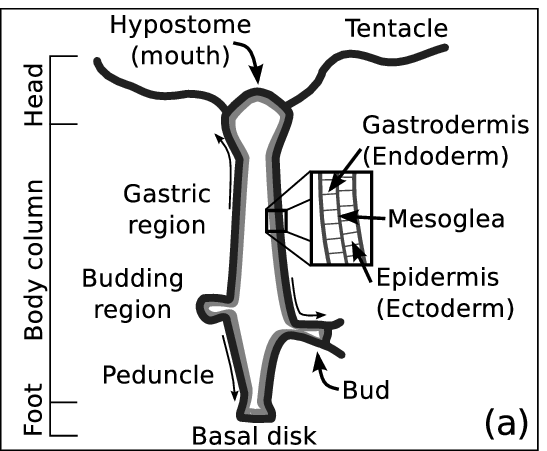}
\includegraphics[height=4.5cm]{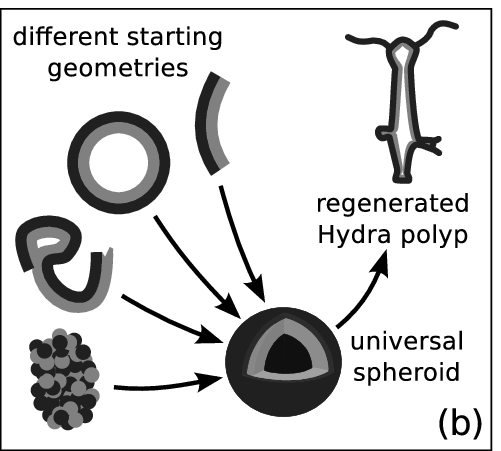}
\includegraphics[height=4.5cm]{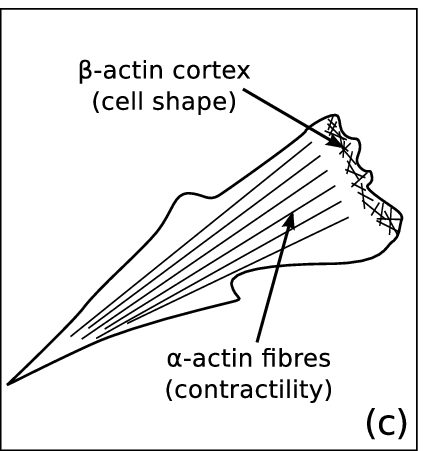}
\caption{(a) Cross-section of a Hydra polyp with two buds. Hydra consists of two
cell layers, the gastrodermis and the epidermis (also called endoderm and
ectoderm), attached to an extracellular matrix called mesoglea. The polyp forms
a tube of which one end is surrounded by 7--12 tentacles with the hypostome
(mouth) in the centre while the other end (basal disk) is used to attach to
surfaces. (b) Tissue fragments and cellular aggregates of different
geometries first transform into the universal spheroidal geometry prior to the
regeneration of a new polyp. (c) Hydra possesses two different isoforms of
actin,
$\alpha$-actin which can build up super-cellular structures, and $\beta$-actin
which becomes particularly prominent when the cell starts migrating out of the
tissue collective (modified from \cite{Gunning1997a}).}
\label{hydra_hydra_sketch_and_actin_types}
\end{figure}

Our multi-cellular system of choice is the cnidarian \textit{Hydra vulgaris}.
It displays a simple and uniform morphology (see
\fref{hydra_hydra_sketch_and_actin_types}~(a)) and possesses only a small
number of cell types. In contrast to many other multi-cellular organisms,
signs of ageing could not be stated, so
``eternal life'' was accorded to this organism \cite{Martinez1998}. Its 
reproduction and regeneration capabilities are stunning:
Hydra cell assemblies and fragments prove to survive and even regenerate
completely. The absence of tissue degradation and decomposition avoids
misleading
results.
These properties, together with the fast proliferation, render Hydra
an ideal model organism for research on bio-mechanics and pattern formation
in tissues. 

Hydra has inspired Alan Turing to his seminal reaction-diffusion principle and,
indeed,
numerous grafting experiments \cite{Wetzel1895,Hefferan1901,Mutz1930} could
be interpreted by postulating local activator and global inhibitor gradients
as proposed by him \cite{Turing1952} and elaborated by Gierer \&
Meinhardt \cite{Gierer1972a,Meinhardt2000a}. Despite great success (e.~g.
explanation of the existence of a minimal tissue size for regeneration), the
gradient-forming molecules still have not been clearly identified
\cite{Bode2009}.
Further, a diffusion mechanism across or outside of
the tissue as required for building such gradients would hardly be precise and
stable enough to control the observed patterning. Unfortunately, Turing did 
not take into account any cell-mechanical aspects, though,
regenerating Hydra tissues, as well as other tissues,
show distinct active mechanical movements. As a
conclusion we hypothesize that forces and movements  are a crucial component
for a stable regeneration of the organism.

It was shown that mechanical stress -- under certain conditions --
influences the chemical state of cells, e.~g. $\beta$-catenin increases
significantly on compression. Furthermore, $\beta$-catenin not only influences
the regulation of the cytoskeleton but also the expression of genes well-known
from
development and cancer \cite{Whitehead2008,Fletcher2010,Farge2011}. However, the
link to the tissue fluctuations and movements is still to be explored.

Single cells revealed singular material properties, partially due to
their highly dynamic polymer networks. The cytoskeleton built out of these
polymers shows complex rheology partially depending on the mechanical past of the
cell \cite{Fernandez2008,Trepat2007}.
This can theoretically be captured since only recently
\cite{LarsWolff2010,Janmey2007,Wen2011}. Furthermore, the cell reacts
specifically to mechanical stress with softening or stiffening dependent on the
entanglement of the fibres and the time scale of observation
(``stiffening-softening paradox``) \cite{Wang2002,Park2010,Wolff2012}.

In most healthy grown tissues cells usually neither strongly change shape nor
migrate. However, it has been stated that tissue grafts lead to increased local
cell motility \cite{Fujisawa1990} and developmental gene activation (Wnt)
\cite{Chera2011} in Hydra organisms. In regenerating tissues cells equally show
increased motility and Wnt-activity similar to single cells
\cite{Hobmayer2000,Galliot2010}.
It is plausible that this developmental gene may be related to cell motility and
healing. Its relation to our findings remains to be studied.

What determines the large scale ordering during regeneration and development?
One mechanism was found by Johannes Holtfreter who investigated embryonic
tissues and suggested cell-cell affinity as a sorting mechanism
\cite{Holtfreter1939}. Foty \& Steinberg showed the direct dependence of surface
tension on adhesion strength between cells in cellular aggregates
(``differential adhesion hypothesis'') analogous to demixing of immiscible
fluids \cite{Foty2004,Foty2005}. Cell assemblies represent a unique material
being able to switch between fluid-like, solid-like or a material with
mixed properties.
Hydra tissues, as studied here, are extracted from adult animals and possess an
extracellular matrix and stable inter-cellular junctions. In contrast to
embryonic cell assemblies they rather behave like a soft solid material.

Fluctuations during Hydra regeneration have been investigated only rudimentarily
so far \cite{Fuetterer2003,Soriano2006,Kosevich2006} and only few publications
discuss fluctuations during morphogenesis for other species
\cite{Koth2011,Solon2009}. It was found that tissue fragments and cellular
aggregates always rearrange to spheroids
(\fref{hydra_hydra_sketch_and_actin_types}~(b)). These spheroids show three
phases
of sawtooth-like semi-periodic fluctuations \cite{Fuetterer2003}. These phases
were found to be related to the expression pattern of a gene associated to the
mechanical axis formation \cite{Soriano2009}. Fluctuations may directly be
coupled
to gene expression, however, many open questions remain.

In order to measure macroscopic shape changes with a high signal-to-noise ratio
GFP-labelled cells have been observed.
Therefore we used strains with fluorescent eGFP being co-expressed along with
$\beta$-actin in the epidermal as well as in the gastrodermal cells, and we
studied them by confocal microscopy. As this isoform was found to be
uniformly expressed, we used the variations in fluorescence intensity as
an indicator for the deviation from the focal plane caused by tissue
deformation.

Hydra cells also possess muscular $\alpha$-actin forming
myoneme-like, force-generating structures whereas cortical $\beta$-actin is
rather involved in the control of stiffness and shape.
Both systems are stabilized and dynamically restructured by
motor proteins (myosins) and crosslinkers (e.~g. actinin) \cite{Gunning1997a}.
However, we ignore the dynamics on a molecular scale, but concentrate our
discussion on the principal functional subsystems: the mesoscopic filamentous
structures denoted as ``$\alpha$-actin bundles/myonemes'' and the ``cortical
$\beta$-actin''.

While we concentrate on the mechanical properties here, it is clear
that the ``big picture'' has to associate mechanics with signalling and genetic
control. We expect that our findings are of general importance for biological
pattern formation, complex systems and may lead to the unfolding of new medical
approaches.

\section{Folding dynamics}

Fragments of different shapes were found to reshape into a spheroid in over
90\,\% of cases. The tissue often rejects a larger number of cells during this
folding process. The passage through the
spheroidal state has been found without exception prior to the regeneration of a
polyp, however, the reason of this necessity is not clear.

\begin{figure}[htp]
\centering
\includegraphics[width=\textwidth]{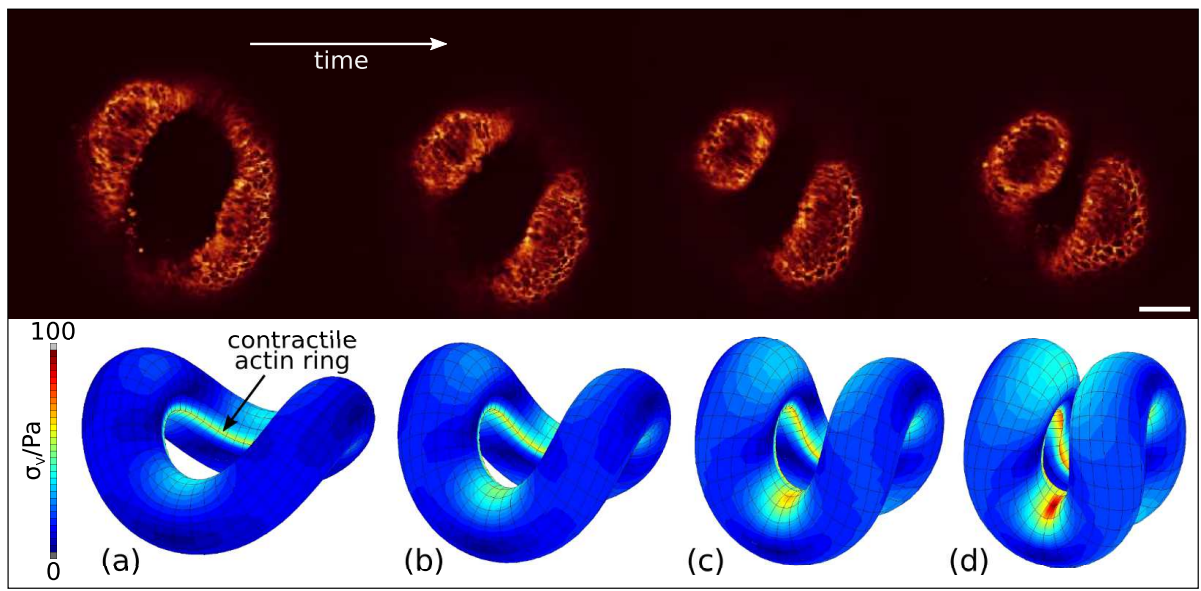}
\caption{(a--d) The figure shows the Hydra folding process (bar: 100\,\textmu m)
observed with confocal microscopy (upper row) and as a simulation (lower row).
The confocal images only show a horizontal cross-section of the 3D-structure. In
the results of the simulation (discussed in \sref{simulation}) the location of
the inner actin ring
(toroid diameter 300\,\textmu m, cross section diameter 90 \,\textmu m) is
indicated by the increase of stress along the inner circumference of the torus
due to myosin-actin contraction. The vertically arranged states correspond
approximately.}
\label{hydra_folding_mechanism}
\end{figure}

In order to obtain uniform and comparable temporal regeneration dynamics we used
toroids as an initial state (\fref{hydra_folding_mechanism}). This simple shape
mimicks an infinite tissue for signal spreading and facilitates data analysis and
the building of models.
The dimensions of our toroidal cross-section are about $80
\times 140$\,\textmu m (radial $\times$ coaxial direction) and
300\,\textmu m (overall diameter).
The toroid's wall consists of a massive inner (gastrodermis)
and shell-like outer (epidermis) cell layer.
The toroids comprise about
1500$\pm$500 cells in total and for this arrangement we found the regeneration
to a small polyp being reproducible in about 80\,\% of our experiments.
In the remaining cases we did not observe folding.
Instead the tissue just contracted until the inner aperture was closed or
the toroid disintegrated completely.
In the case of too small sections the folding still occurs but the reproduction
probability of the polyp is reduced.
Below sizes of 200-300 cells the regeneration fails \cite{Shimizu1993}. 
Too large sections do not fold but stay tube-like and heal at both ends prior
to regeneration. In that case the axis of the organism is presumably conserved.

The folding process in 90\,\% of our observations requires not more than
$(120\pm30)$\,s from the planar ring-shape to the folded ring.
The folding was considered as completed when the opposite loops got
into contact. 
The observed time period is clearly too short for diffusive signalling
across
the toroid, especially as an appropriate control loop would need several
passages of wave fronts of signalling molecules before a gradient obtains
stability. Half of the perimeter accounts for at least 20 cells and free
diffusion would disperse a signal in not less than 10\,min to reach the opposite
side \cite{Francis1997}.

During wing morphogenesis of the fruit-fly a Dpp (morphogen)
gradient expansion speed of 6\,\textmu m in 5~hours (this corresponds
to 3~days to cross a Hydra toroid) has been measured \cite{Entchev2000},
which is by far too slow to explain Hydra toroid folding.
Gene expression would also need
many hours \cite{Noble2010,Cheadle2005,Fan2010}.
A sufficient control of diffusion based on gradients outside of the tissue is
hardly imaginable. In addition, the Hydra polyp lives in an aqueous environment
which would strongly perturb such gradients.

Hydra possesses
a primitive neuronal system mainly concentrated in the hypostome and peduncle
region \cite{Grimmelikhuijzen1982}. The toroids are taken from the centre of the
gastric column which is only sparsely populated with neurons. As
most of their connections are destroyed
during the dissection process, we assume that their contribution to the
control of the folding process is at best
marginal.

Other signal paths are provided by gap-junctions, prominent for cardiomyocytes
but still unknown for Hydra.
They allow for a direct and extremely fast intercellular signal exchange based
on electrical potential differences driving ion flow
\cite{McDowall1980}. However, an organizer as the sinoatrial node for the heart
would be required to provide timing stability. Such a system is unknown in Hydra
and presumably negligible in our toroids.

Osmotic pressure of the gastrodermal cells as a possible origin of
contraction and deformation can be excluded since the enteron (inner cavity of
a closed Hydra tissue) is hyperosmotic. This would result in a cellular swelling
and not a contraction once these cells are exposed to the external medium
\cite{Kuecken2008}.
These reasons support our conclusion that the gastrodermal cells
are the force-generating cell type.

Mechanical stress-relaxation waves, in contrast, propagate at the speed of
sound and provide a means of very fast signal transmission. The
corresponding speed $v = \sqrt{G/\rho}$ is in the range of about 0.1\,m/s
when assuming a shear modulus $G \approx 100\,$Pa (soft cells; stiffer cells
lead to even higher velocities) and a net tissue density $\rho \approx 1$\,g/ml
\cite{OrescaninM2009}.
Mechanical waves cross Hydra rings in milli seconds.
The shear modulus is controlled by the cell cortex, which stabilizes
cellular shape against external mechanical stress and osmotic pressure
\cite{Kuecken2008}.

During the folding process, the gastrodermal cells in the fold are submitted to
a considerable compression leading to strong deformation. In some cases
this deformation results in a local tissue disassembly as some cells start
migrating individually first, then they round up, their $\beta$-actin related
fluorescence is strongly increased and, finally, some quit the tissue. This
process resembles the epithelial-mesenchymal transition (EMT) which plays a role
in tumours and inflammation, for stem cells and during embryogenesis
\cite{Zhao2011,Dave2012,Holley2007}. To our knowledge, a purely
mechanical triggering of this transition has not been described before.

Cells remaining tissue-bound show a low, constant and uniform $\beta$-actin
activity. The cortex provides stiffness to assure the stability of the
cells and the tissue.
Even for strongly deformed cells an increase of the corresponding fluorescence
intensity could not be stated as long as the cells remain tissue-bound.
These observations agree with gene expression studies where the
$\beta$-actin expression rate has been found stable enough to
serve as a reference for normalization of gene expression measurements
\cite{Ferguson2005}. However, this statement has to be revised in our case as we
observed significantly higher activity of $\beta$-actin once the cells switch
from the “tissue state” to the individual migrating state.
As we did only observe exclusively tissue-bound (low fluorescence) or migrating
(strong fluorescence) cells, we suggest a two state approach for future models.


\section{The actin machinery}

The $\alpha$-actin system of Hydra forms super-cellular bundles in the epidermis
(myonemes) as well as in the gastrodermis. They are able to span across as much
as 7 cells. One epidermal cell contains about 7--10 bundles. The bundles in the
two cell layers are oriented orthogonally to each other and form a
two-dimensional cartesian coordinate system, which allows to absorb as well as
generate stress in any direction. This explains the impressive motility of the
organism. The epidermal bundles are oriented coaxially to the Hydra body and
the dissected-toroid axis, and they are positioned regularly with an average
distance of 3--5\,\textmu m. The gastrodermal bundles follow the contour of the
toroid, with strongly varying density. We observed strong bending and length
fluctuations in both systems. The gastrodermal bundles are much less pronounced
than the epidermal bundles and usually appear more clearly once the tissue is
slightly stimulated mechanically.

\begin{figure}[htp]
\centering
\includegraphics[height=4.5cm]{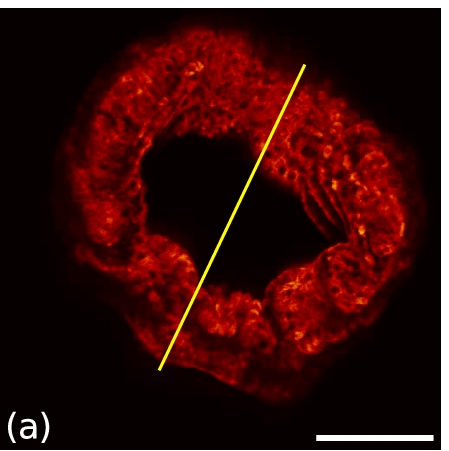}
\includegraphics[height=4.5cm]{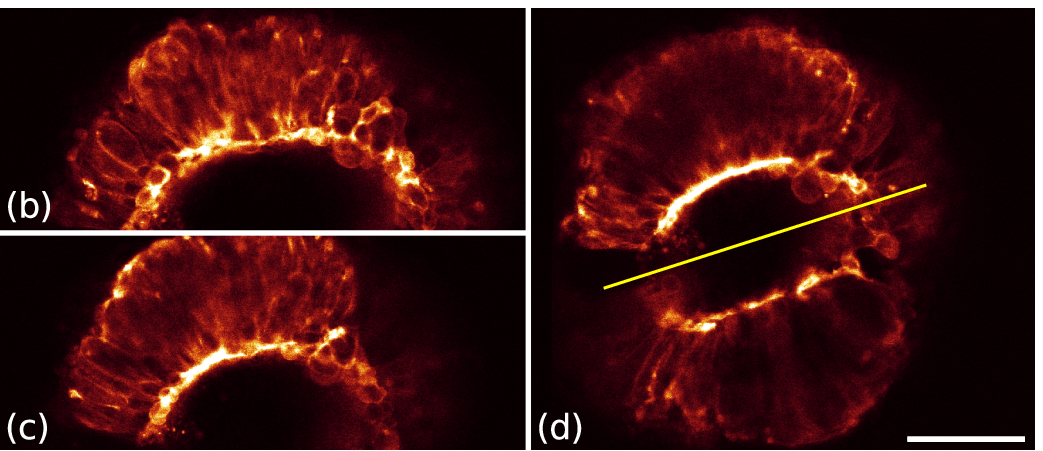}
\caption{(a) The $\beta$-actin fluorescence intensity distribution in the
gastrodermis of a projection of a partially folded (about 50\,\%) Hydra does not
indicate the axis along which the folding occurs later (axis in yellow).
Therefore we think that the folding axis selection is random. (b--d) shows a
sequence of gastrodermal actin bundle formation in a later state of the folding
process. Initially, (b), the actin is scattered over the apical sides of the
cells. After about 2 1/2 min, (c) the bundle starts forming and at later times
(d) it becomes straight and dense (folding axis in yellow). All bars represent
100\,\textmu m.}
\label{endo_la_bundle_formation}
\end{figure}

In \fref{endo_la_bundle_formation}~(a) the toroid just
started the folding process. The observed $\beta$-actin fluorescence intensity
did not display any specific dynamics during that process neither in the
gastrodermis
nor in the epidermis. We conclude that $\beta$-actin may rather serve for
maintaining a uniform stiffness of the cellular material.
The fluorescent gastrodermal actin forms bright zones prior to the
folding event. Initially the actin is scattered in the apical cortex of the
irregularly shaped gastrodermal cells. In course of time the actin structures
become more dense and get aligned to bundles
(\fref{endo_la_bundle_formation}~(b--d)). Finally, a dense and strong actin ring
is formed along the inner side of the toroid and the cell's apical side is
flattened to a smooth inner contour. This is presumably due to increasing
internal mechanical stress reducing the surface roughness. It is conceivable
that the bundling process itself is self-sustained and amplified by this stress
along the curved geometry. Simultaneous to the bundle formation we observed a
decrease in fluorescence intensity of the cytoplasm probably due to actin
depletion.

The epidermis arches as a relatively thin layer over the outer bound of the
gastrodermis, which is much more voluminous. Due to their orientation, the
epidermal $\alpha$-actin bundles cannot be directly responsible for the
folding. We assume that one of
their duties is rather to distribute the stress field generated by the
contracting gastrodermal bundle ring over the entire toroid. This assures
stability and reproducibility of the described dynamics.

\begin{figure}[htp]
\centering
\includegraphics[width=\textwidth]{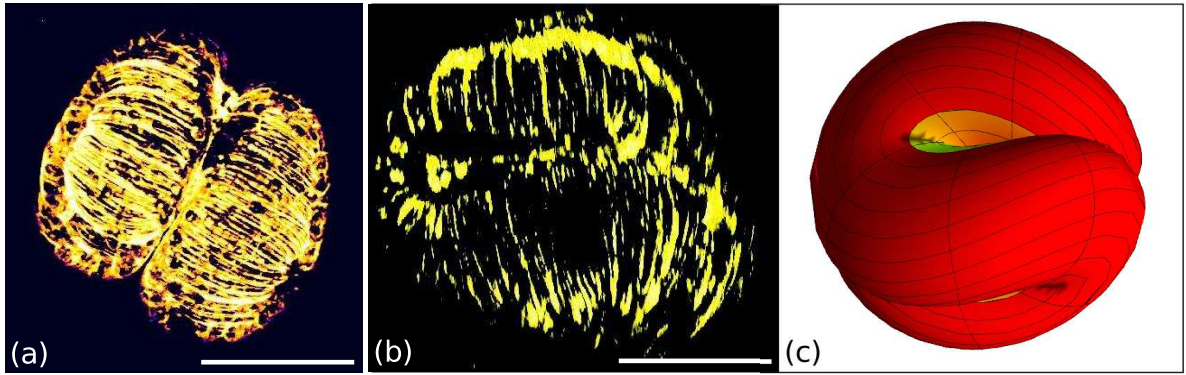}
\caption{(a) top view of the epidermal $\alpha$-actin structure (myonemes) which
(b) builds arches over the gastrodermal loops (bars: 100\,\textmu m). The
stiffness of these bundles stabilizes while providing enough flexibility to
follow the transverse and longitudinal fluctuations due to the gastrodermal
cells. The orientation of (b) is visualized in (c).}
\label{hydra_ecto_lifeact_structure}
\end{figure}

The epidermis covers as a relatively thin layer the voluminous gastrodermis
and possesses a system of long and equidistant epidermal bundles (\fref{hydra_ecto_lifeact_structure}).
Their length was observed to fluctuate between 10 and
80\,\textmu m with rates up to 150\,\textmu m/min. Actin polymerization is
clearly too slow to yield such rates, hence myosin is assumed to be at the
origin \cite{Kuhn2005}. The gastrodermal tissue beneath is periodically
compressed by these fluctuations which would explain the observed densification
and orientation of the gastrodermal actin structure. These contractile
forces are transmitted to the adjacent cell layer by the flexible and porous
extracellular matrix
network \cite{Shimizu2008}. The a priori highly oriented epidermal
bundles presumably determine the orientation of the gastrodermal bundles which
in turn generate the mechanical stress expressed in transversal epidermal
fluctuations.
The gastrodermal system was observed to regularly fractionate again and split up
between the epidermal contractions. So it is much less stable than the epidermal
one which may allow it to be more adaptive with respect to external changes in
stress and shape.

\begin{figure}[htp]
\centering
\includegraphics[height=4.5cm]{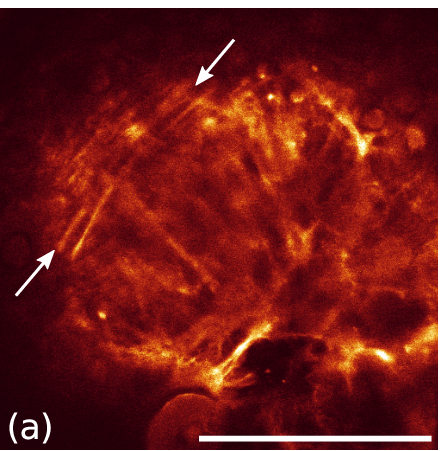}
\includegraphics[height=4.5cm]{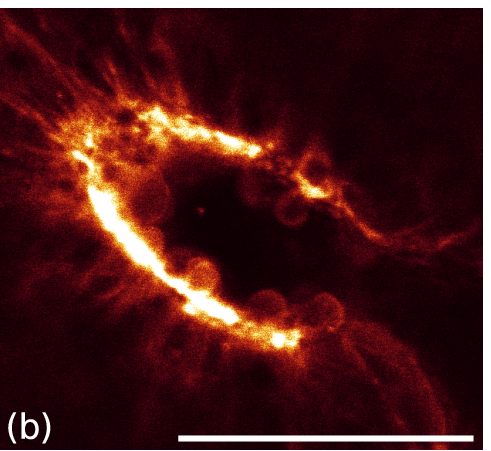}
\includegraphics[height=4.5cm]{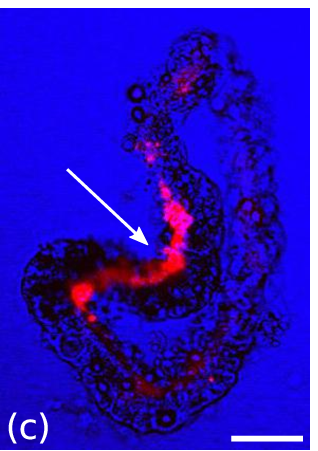}
\caption{(a) The gastrodermal $\alpha$-actin bundles usually are faint. However,
if the fast deformation leads to an internal shear stress these structures are
expressed more strongly (ends of bundles indicated by arrows). The folding axis
is oriented horizontally. (b) During folding, the actin intensity is strongly
increased on the apical side of the gastrodermal cells, indicating the
contraction of this cell layer. (c) Cytochalasin at concentrations above
20\,\textmu mol/$\ell$ destroys the gastrodermis whereas the epidermis seems to
be more stable. Still some gastrodermal cells (red) remained intact in this
picture. The curvature of the ring is more pronounced at that sites (indicated
by the arrow). This shows the crucial role of the gastrodermis for the folding
process. All bars represent 100\,\textmu m.}
\label{hydra_folding_images}
\end{figure}

\begin{figure}[htp]
\centering
\includegraphics[width=\textwidth]{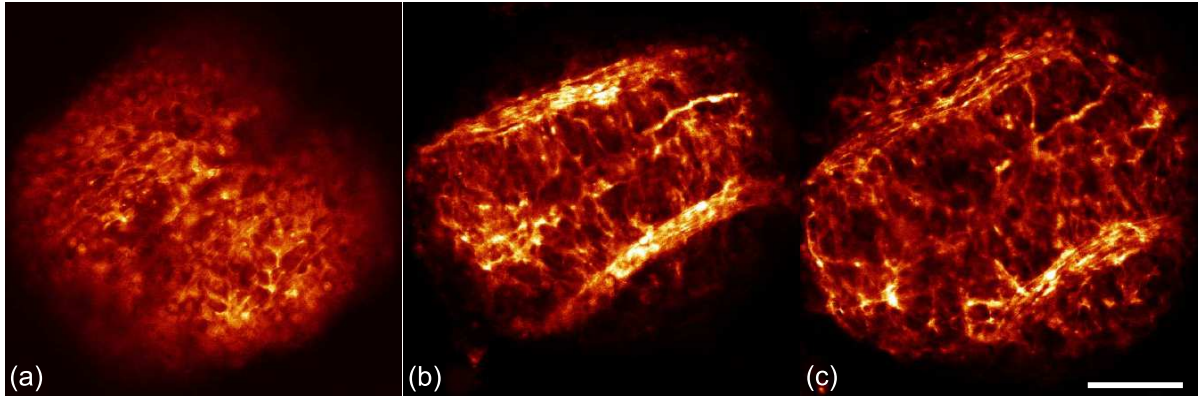}
\caption{The actin ring seems to play an important role even in the already
folded ring. Slight compression significantly amplifies this otherwise only
hardly visible structure (a). The cells of the epidermis and the gastrodermis
are pressed together and after 1/2 hour (b) the cells in contact connect
inducing the closure of the gaps. Finally (c), the $\alpha$-actin bundles start
to disappear and a perfect spherical symmetry is established (bar: 100\,\textmu
m). This spheroid, being symmetric in shape and mechanical properties (actin),
is the starting point inevitable for the development of a novel organism.}
\label{hydra_hydra_under_glass}
\end{figure}

We hypothesize that the gastrodermal actin ring as seen in
\fref{endo_la_bundle_formation} (b--d) and in 
\fref{hydra_hydra_under_glass}
(b--d) is responsible for the folding process.
This is supported by partially dissolving the gastrodermis by application of
cytochalasin --- an actin polymerization inhibitor. Degradation of the
gastrodermis results when doses above 20\,\textmu mol/$\ell$ are applied for 10
min. The epidermis is less prone to degradation than the gastrodermis. In
\fref{hydra_folding_images}\,(c) it can be observed that the epidermis is
significantly more curved in regions where some gastrodermal cells are still
attached to the tissue.

\section{Differential contraction and toroidal symmetry breaking}
\label{contraction}

In order to perform the described folding process the cylindrical symmetry of
the toroid has to be violated. Due to the contraction of the actin ring in the
gastrodermis the whole tissue experiences an internal stress gradient
(``differential contraction``) between the actin-ring-forming and the other
cells and the toroidal shape becomes unstable. Small randomly distributed
irregularities (''critical fluctuations``) may be amplified now.
As a consequence the tissue increases its
curvature transversally and becomes wavy. The nature of the irregularities is
not obvious, as thermal fluctuations are negligible at this length scale. The
origin of these active fluctuations is presumably linked to the actin
cytoskeleton which is known to be highly dynamic and a source of
fluctuations \cite{Wen2011}.

In this section we relate single cell mechanics to the described fluctuations
of Hydra tissue toroids.
The mechanics of single cells under different types of external forces
and strains is currently being investigated extensively
\cite{Trepat2007,Angelini2010,Lin2010,Klein2009,Janmey2007,Tee2009,Tee2011}.
The spatial scale of these fluctuations was found to be larger
than just a single cell, therefore, it presents a collective phenomenon.

During folding cells are deformed strongly.
This can be lead to disassembly of the cytoskeletal actin crosslinkers.
These crosslinkers are point-like and, therefore, they concentrate the
mechanical stress field strongly which increases rupture probability.
Therefore even small strains ($> 10 \%$) lead to an irreversible actin network rupture \cite{Wolff2012,Trepat2007}.
The strain-softened cells extend and the overall stress is relaxed.
However, the disrupted cytoskeletal structure of these cells
reorganize an stiffen slowly again after several minutes \cite{Trepat2007}.
As the recovered stiffness exceeds the stiffness of the non-softened adjacent cells 
as can be seen in \cite{Trepat2007},
the latter are stretched and shear-softened during a new folding trial.
Therefore repeated folding would occur along a varying axis.
Indeed, we occasionally observed toroids to unfold and refold at a different
axes.
Apparently the toroids ''check out`` if the folding was correct
and repeat it on missmatch.

Cells actively react on stress. In preliminary experiments with toroids exposed
to strong mechanical stress (2--5\,\textmu N) in a mechanical stretching
device we were unable
to predict the position of rupture. $\alpha$-actin was found to reinforce by
bundling at the thinnest site presumably permitting the tissue to
cope with the densified mechanical stress field. The active
reinforcement of actin bundles in the gastrodermis protects the tissue from
rupture.

\begin{figure}[htp]
\centering
\includegraphics{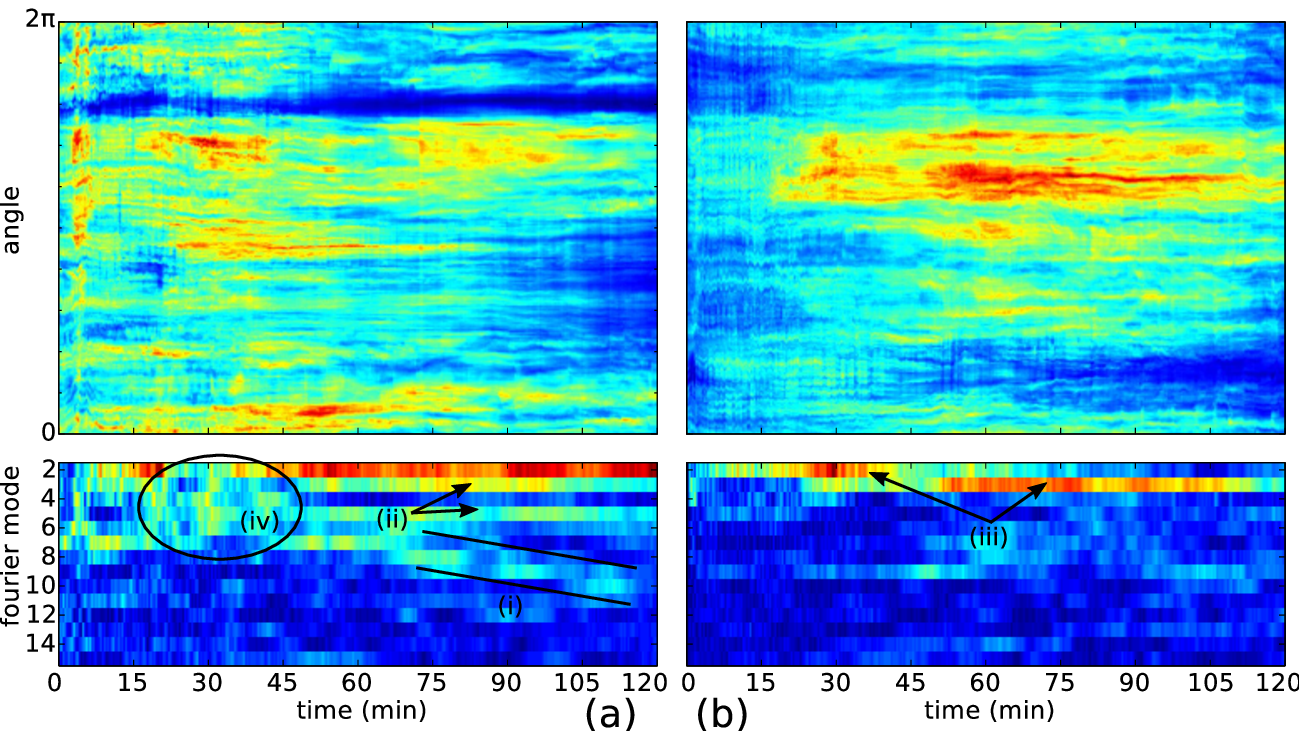}
\includegraphics{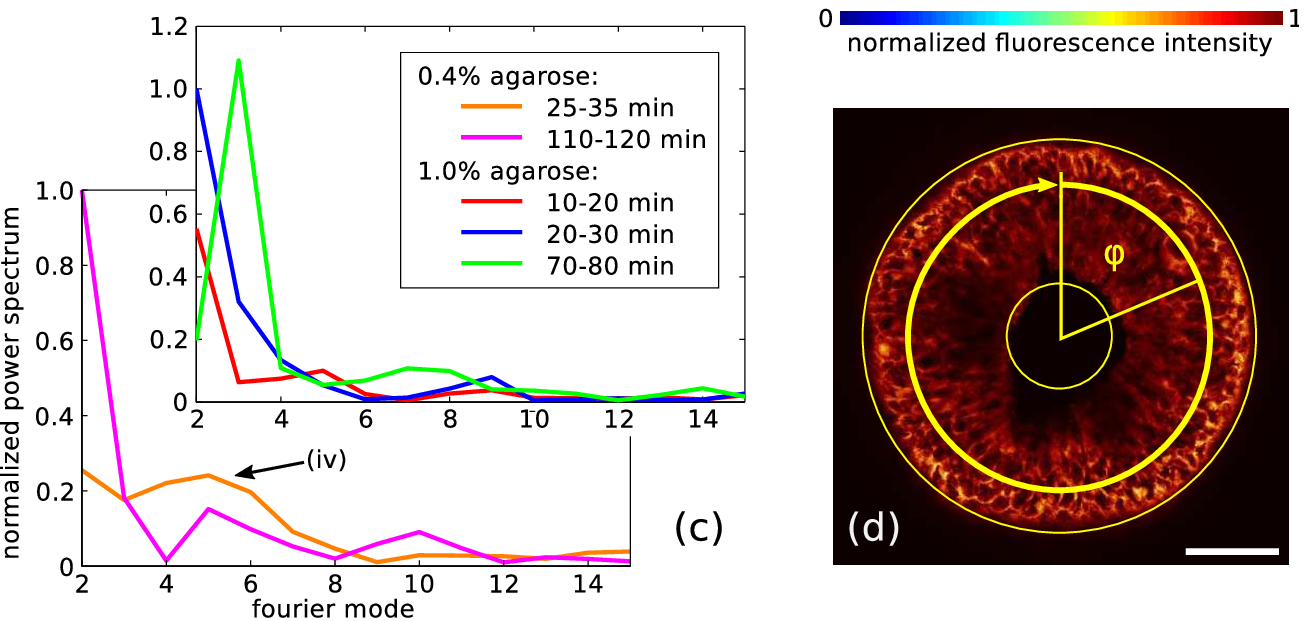}
\caption{Hydra tissue in low melting agarose gel with concentrations of (a) 0.4,
and (b) 1\,\% is shown here.
The gel inhibits the folding process and allows the long-term
observation of the mode dynamics on the toroid. (a) Previous to the folding we
found presence of the modes 2--7 with similar amplitudes. After a few minutes
the higher modes disappear in favour of the 2\textsuperscript{nd} mode finally
leading to the folding process. We found a cascade dissipation mechanism (i) as
well as the coupling of a number of even or odd modes (ii) reflecting even or
odd mirror symmetry. In (b) the very stiff gel results in a winning
3\textsuperscript{rd} mode (iii). No higher modes are significant here. (c)
Spectra, normalized with the initial value of the 2\textsuperscript{nd} mode and
averaged over a short interval at the indicated times, are compared for the two
gels: In the softer gel a block (2-6) of modes are of equal strength (iv) during
the symmetry breaking (25-35 min), which decay later. Only the
2\textsuperscript{nd} mode survives and dominates finally. In stiffer gels no
block could be seen, usually the 2\textsuperscript{nd} mode dominates during the
transition. The presented case was observed in the stiffest gel: mode switching
from the 2\textsuperscript{nd} to the 3\textsuperscript{rd} was found. The polyp
was not able to be regenerated in this case. (d) A typical tissue ring with
GFP-labelled epidermis is shown together with the sampling strip along which the
intensity was extracted and radially averaged for the Fourier analysis.
(bar: 100\,\textmu m)}
\label{hydra_waves_on_the_ring}
\end{figure}

The tissue fragment folds rapidly. When embedded in very soft agarose gel of
concentrations from 0.2 to 1\,\% \cite{Normand2000}, the folding onset can be
retarded or stopped allowing for longer observation times. On a long term
(about 1 hour) we found three phases of shape fluctuations -- first a
semi-periodic phase with typical frequencies in the range of 10\,mHz, then a
second phase with pulsations every few minutes -- and finally a silent phase. In
the last phase the tissue organization starts to disintegrate partially (EMT).

Regarding the initial fluctuations leading to the instability, we observed
mainly creation and decay of stationary waves.
Corresponding to the periodicity of the system we used discrete Fourier analysis
of the fluorescence intensity along the toroid with the toroidal angle as
variable. We restricted our analysis to the modes 2 to 15. Higher modes would
account for sub-cellular deformations which go beyond the scope of this
publication. Modes 0 and 1 correspond to translation and rotation and are
therefore irrelevant for the folding dynamics.

Our data are discussed qualitatively only as the described
phenomena are reproducible, though, not yet numerically.
Initially, several of the lowest modes (2--10) were of about equal
amplitude (see (iv) in \fref{hydra_waves_on_the_ring} (c)). At the time scale of
several
10\textsuperscript{ths} of minutes, all modes decayed with exception of the
2\textsuperscript{nd}. This mode led directly to the correct folding geometry.
For stiffer gels we observed a reduction of excited modes and a slowing down of
the dynamics. In an almost liquid 0.2\,\% gel the second mode dominated
after less than 5 min, in stiffer gels it needed significantly more time. Only
in the stiffest gel (1\,\%) the 3\textsuperscript{rd} mode was able to supersede
the 2\textsuperscript{nd} in the end (see (iii)
in \fref{hydra_waves_on_the_ring} (b)).
This mode exchange can be explained by considering the distribution of
the mechanical energy.
We consider bending into the direction of the toroidal axis only and neglect
modulations in the toroid plane.
The bending energy of the toroid scales for excursion amplitudes $a$, which is 
small, like $E_\mathrm{bend} \sim a^2\, n^4$ ($n$ is the mode number).
In the gel-less case the energy is distributed
equally among the modes according to the equipartition theorem.
Then, the lowest modes dominate since $a \sim 1/n^2$.
In linear approximation and assuming that the average force applied against
the gel $F$ is constant, the elastic energy of the gel is $E_\mathrm{el} \sim
F^2/D$ ($D$ is the elasticity constant)
and the energy created by the contraction of the actin ring
increases in time. The contraction process continuously delivers mechanical
energy into the system being distributed between the gel and the bent toroid.
However, the stiffer the gel is, the less energy it can store: $E_\mathrm{el}
\sim 1/D$. Therefore, for stiffer gels the energy
generated by the contraction goes preferably
into the tissue deformation. Eventually, the even 2\textsuperscript{nd} mode is
not absorptive enough anymore and the odd 3\textsuperscript{rd} is involved to
take
over the excess energy. As the latter can store
$5\times$ more energy compared to the 2\textsuperscript{nd} mode at equal
excursion
amplitudes ($E_\mathrm{bend} \sim n^4$) and dominates and suppresses the
2\textsuperscript{nd} mode by still unknown non-linear mode coupling.

The modes superior to the 2\textsuperscript{nd} one frequently decayed in a
cascade through which their energy was progressively transferred to increasingly
higher modes (a typical case is shown in (i) in \fref{hydra_waves_on_the_ring}
(a)).
This again can be explained by the better ability of higher modes to absorb the
increasing amount of mechanical energy generated by the contracting actin ring.
This might be a biological dissipation mechanism to transfer the steadily
increasing energy from macroscopic to mesoscopic and possibly
microscopic length scales, i.~e. to the molecular level.
The energy is completely transferred to the next higher mode, which indicates
again a non-linear competition of modes with different symmetries.
We generally observed transient coupling of exclusively odd or even modes
(a typical case is shown in (ii) in \fref{hydra_waves_on_the_ring} (a)).
The modes of equal symmetries collaborate and modes of mixed symmetry
compete.
However, an explanation is still unavailable.

Finally, after a longer period when the folding process failed, cells
round up, increase $\beta$-actin expression, form lamellipods and start
migrating individually over the remaining tissue. We assume to have
observed for the first time a purely mechanically triggered
epithelial-mesenchymal transition.

\section{Finite element simulations}
\label{simulation}

In this section we describe numerical simulations of the folding process
using a three dimensional finite element model. The calculation assumes
quasi-equilibrium and outputs the state as a function of the stress generated
by the contractile actin ring.

In order to account for large deformations, an \textit{Updated Lagrangian
formulation} \cite{Belytschko2000} is chosen. The deformation behaviour is
modelled by an Ogden material model of isotropic non-linear elasticity
\cite{OGDEN1972,OGDEN1972a,Simo1991}, characterized by the free energy function 
$\psi\left(\lambda_k\right)=\sum_{I=1}^n
\frac{\mu_I}{\alpha_I}\left(J^{-\frac{\alpha_I}{3}}\left(\lambda_1^{\alpha_I}
+\lambda_2^{\alpha_I}+\lambda_3^{\alpha_I}\right)-3\right)+g(J),$
with principal stretches $\lambda_i$, material parameters $\mu_I$ and $\alpha_I$
as well as $n$, the number of individual functions. The function $g(J)$ of the
Jacobian $J=\lambda_1\lambda_2\lambda_3$ is used to model compressible material
behaviour. Here, we use an Ogden formulation with $n=1$, $\alpha_1=2$ and
$g(J)=\frac{9K}{2}\left(J^\frac{1}{3}-1\right)^2$, which is also known as a
compressible Neo-Hooke material, where $K$ is the bulk modulus. In the limit
case of small strains this formulation reduces to linear elastic Hooke material.
Regarding the material parameters, we chose a Young's modulus of $E=100$\,Pa and
a Poisson's ratio of $\nu=0.4$. The Poisson's ratio quantifies the negative
ratio of transverse and longitudinal strain in a specimen undergoing uniaxial
tension. The chosen number 0.4 allows for a small volume increase on extension,
meaning the material is assumed to be slightly compressible. From those
parameters $\mu_1=\frac{E}{2(1+\nu)}$ and $K=\frac{E}{3(1-2\nu)}$ can be
calculated.

Tori with major radii of $R=150$\,\textmu m and varying minor radii $r$ have
been investigated. These were discretized by \textit{hexahedral serendipity
elements} with quadratic shape functions \cite{Cook2001}. The inner actin ring,
assumed to be responsible for the folding process, was modelled by linear truss
elements. These are attached to the toroid along its inner circumference. To
drive the folding process an increasing intrinsic strain was prescribed to the
truss elements.

Simulations showed that numerical noise is not sufficient to break the symmetry
of the toroid model. Thus four equal additional forces distributed evenly around
the toroid are applied, forcing the toroid slightly into the experimentally observed
configuration. While reducing these additional forces back to zero, the
simultaneously increasing intrinsic strain in the inner actin ring will keep the
toroid in its bended shape. Further increase of the intrinsic strain then drives
the folding. The hereby described process is adequate to prove the ability of
the inner actin ring to fold the toroid if the initial condition describes a
sufficiently bended configuration.
We suggest that the active fluctuations described in
\sref{contraction} serve to overcome this folding threshold.

\Fref{hydra_folding_mechanism} features a toroid of 45\,\textmu m minor radius
modelled by 2304 hexahedral and 96 truss elements. As the simulations show, the
inner actin ring is able to fold the toroid, which proves the viability of our
hypothesis. In \fref{hydra_folding_mechanism}~(d) the inner ring exhibits
tensile forces between about 50 and 150\,nN. This results in von Mises stress
$\sigma_\mathrm V$ of up to about 100\,Pa in the toroid. Thicker toroids did not
succeed
in folding but would rather return to their plane configuration when reducing
the additional forces. This is to be expected since for thick structures bending
becomes less favourable compared to tension. Therefore, simulated toroids with
aspect ratios $r/R > 0.3$ would rather tighten staying flat than to deflect from
of their plane configuration which we occasionally could also observe in our
experiments.

The model is well suited to describe the observations qualitatively, whereas any
numerical quantity values should be regarded as describing the order of
magnitude of effects. Simplifications of the model are: ignoring the nature of
the two cell layers as well as the presence of the extracellular matrix
(mesoglea), the other actin structures, and finally, the simulation is based on
a simple isotropic Ogden model. Furthermore, apart from the contractile actin
ring we ignored active cellular reactions which modulate cellular stiffness and
shape and which are presumably responsible for the described fluctuations.

\section{Conclusion}

During regeneration, cellular toroids composed of about 1500 Hydra vulgaris
cells display a highly symmetric and unusually fast folding
dynamics.
At the end a compact form is achieved which transforms into a spheroid with
a correct cellular bilayered structure.
It must be emphasized that this process
evolves by far too fast for biochemical signalling and gene expression.
Furthermore, a central organizer as e. g. required for the synchroneous heart
contraction has not been found.
Therefore we assume mechanical signalling to accomplish this transformation.

In order to study the onset of the folding we embedded the toroids into gels of varying stiffnesses 
and analyzed the modulation dynamics by circular Fourier decomposition.
For soft gels the 2\textsuperscript{nd} mode prevails which matches perfectly the folding geometry.
For stiff gels we observed a dominant 3\textsuperscript{rd} mode.
Higher modes are capable to accommodate more bending energy for a given
amplitude than lower modes.
As the amplitude is restricted by the gel constraint the energy
distribution is distorded correspondingly.
Energy transfer cascades to higher modes were also found for
modes beyond the 3\textsuperscript{rd} order.
The observed phases of exclusively excited even or odd modes indicate a
symmetry dependent interaction between modes of different order.
We suggest that cells accomplish control of the large scale geometry 
by tissue bending modes.
They also are able to store mechanical energy by this means.
The pulsation dynamics may be required to explore the phase space
for the correct regeneration path.

The subsequent folding process can be explained purely mechanically based on
''Differential Contraction``:
a subgroup of the cells in a tissue contracts collectively which leads to stress
gradients deforming locally the tissue due to its elasticity.
As the driving force of the stress we found the formation of a distinct
mesoscopic contractile actin ring at the inner bound of the gastrodermis.
Latter can lead to the described folding process as we could show by numerical
calculations.
The ring contraction destabilizes increasingly the arrangement of the otherwise
stable flat toroid.
Once a threshold value (found in the numerical model) is reached the tissue
toroid starts bending -- preferably according to the 2\textsuperscript{nd} mode.

Since the $\beta$-actin concentration was not found
to be significantly changed during folding in both cell layers, our observations
lead us to the conclusion that gastrodemal $\alpha$-actin determines the
dynamics.
The differential contraction may self-amplify as the resulting shear stress may
also support the actin filament alignment and bundling.

The finite element simulation revealed that the
ratio between the cross section and the major diameter had to be below a
critical value to accomplish folding. Otherwise the inner bound of the toroid
only contracts without excursion into the third dimension.
Furthermore, it was found that the fluctuations responsible for the transition
have to pass a minimal amplitude in order to initiate the dynamics.
We hypothesize that this is the reason for cells and groups of cells to actively
drive and maintain the observed strong fluctuations in a band of different
modes.

We suggest following mechanical feed-back control loop for the folding:
the epidermis provides equidistant $\alpha$-actin stripes performing
fast longitudinal myosin-driven fluctuations.
They contribute to the bundling and reinforcement of the perpendicularly
oriented gastrodermal actin fibres which finally form the contractile actin
ring.
Gastrodermal contracting fluctuations, in turn, are also fed back to the
epidermal cells,
which results in transverse fluctuations of the epidermal actin structure.
This may present a mechanical closed control loop organizing the described
folding so perfectly.

In the end state of folding the outermost cells join until a double layered
spotless spheroid is obtained.
This is the starting point of the morphogenesis of a new Hydra as described
elsewhere.

\begin{figure}[htp]
\centering
\includegraphics{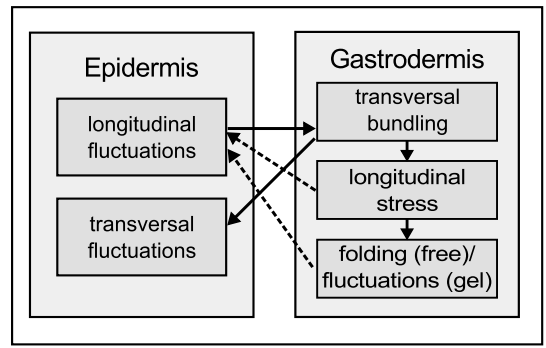}
\caption{The scheme shows the hypothetical control dependencies of the different
actin systems. The fluctuations of the epidermal actin structure bundle the
gastrodermal actin forming a ring. As a consequence the stress is increased
which leads to the folding process or to the transverse epidermal fluctuations
as well as the tissue contractions when embedded in gel. We speculate that 
mechanical feed back (dashed lines) is responsible for synchrony and stability
of the epidermal actin fluctuations.
}
\label{hydra_healing}
\end{figure}

The folding was related to single cell dynamics.
Tissue cells are put under stress by deforming neighbouring cells.
SLarge enough strain stress-softens the actin network which
stabilizes the folding site and reduces the folding force.
The cells remain soft for several minutes after stress release.
Afterwards they recover but the stiffness reinforcement overshoots the
previous value:
in case of faulty folding the tissue changes folding site as a consequence.

The long term observation of tissue toroids in gels revealed a new
mechanically induced transition to individual cellular behaviour (EMT).
The tissue bound cells evade and migrate as individuals over the remaining
tissue.
We also observed the inverse process, i. e. migrating cells
penetrate the tissue again an re-integrate.
The $\beta$-actin level in the migrating state is significantly increased in
contrast to  $\alpha$-actin, which is unexpected.
The cells seem to be able to switch between these two state presumably
corresponding to biochemical signals.
A future theoretical model may therefore be based on two cellular states with
corresponding transition rates depending on external signals.


\section{Materials and methods}

We cultivate four transgenic Hydra vulgaris strains with fluorescence labelled
epithelial-muscle cells either for the gastrodemis or the epidermis. Two
cultures are transfected with the F-actin binding Lifeact peptide
\cite{Riedl2008} whereas the other two cultures express eGFP with a
$\beta$-actin promoter and terminator \cite{Wittlieb2006} simultaneously to
the functional $\beta$-actin of the cells. Therefore, the eGFP signal quantifies
the $\beta$-actin concentration. All strains are kept in crystallizing dishes in
our chemistry lab at temperatures of $(18\pm1)$\,\textcelsius. All cultures are
fed with freshly hatched Artemia salina nauplii once a day and the medium is
changed 3--5 hours after feeding.
Our medium is composed of 1.0\,mmol/$\ell$ CaCl$_2$, 0.1\,mmol/$\ell$ MgCl$_2$,
0.03\,mmol/$\ell$ KNO$_3$, 0.5\,mmol/$\ell$ NaHCO$_3$ and 0.08\,mmol/$\ell$
MgSO$_4$ in Millipore water.

The rings were obtained by dissecting the tissue from the central gastric column
and immediately transferred to a modified petri dish with a 170\,\textmu m cover
slip mounted over an aperture and with a PTFE plate containing holes with a
diameter of 1\,mm. The teflon plate suppresses parasitic convective flow
carrying the Hydra rings out of the observation field. The chamber was filled
either with medium or low temperature melting agarose gel (Sigma-Aldrich A0701)
and all together was completely submerged into Hydra medium to avoid osmotic and
concentration change due to evaporation.

The toroids were observed on a Leica DM IRE2 inverted microscope coupled with a
Leica TCS SP2 AOBS confocal scanner and a Leica HC PL Fluotar 10$\times$/0.30
objective. 

The toroids were made from polyps starved for 24 hours and selected for healthy
shape prior to dissection. A double-blade scalpel was used to cut out the
segments. With this technique we avoid large thickness variations due to polyp
contractions. As the tissue movements are considerable during the first 30\,s
the toroid had to be transferred fast to the observation platform.

The images were visualized and analyzed with ImageJ 1.45s and in-house developed
Mathematica 8.0 and MatLab R2011a scripts.

For the gastrodermal tissue degradation Cytochalasin D (Sigma-Aldrich C8273) was
applied at concentrations up to 20\,\textmu mol/$\ell$ for 10 min. The petri
dish was gently shaken for 10\,s before observation.

The finite element simulations were done using Marc Mentat 2010.1.0. One
calculation for the chosen resolution took about one hour.

\ack
The work presented in this paper was made possible by funding from the German
Federal Ministry of Education and Research (BMBF, PtJ-Bio, 0315883).
We are indepted to Prof. Josef Käs, Prof. Klaus Kroy and Matti Gralka (Leipzig),
Prof. Thomas Bosch and Dr. Konstantin Khalturin (Kiel), Prof. Bert Hobmayer
(Innsbruck), Prof. Albrecht Ott (Saarbrücken), Dr. Roland Aufschnaiter (München)
for many discussions, and providing us with materials, especially the transgenic
Hydra strains as well as giving access to the confocal microscopy post. We are
further grateful for the support of Magna Diagnostics GmbH (Leipzig).

\section*{References}
\bibliographystyle{unsrt}
\bibliography{bibliography}

\end{document}